\documentclass[a4page,12pt]{article}

\usepackage{amssymb}
\usepackage{amsmath}
\usepackage{natbib, graphicx, epsfig}
\usepackage{xcolor}

\newcommand{\be}{\begin{equation}}
\newcommand{\en}{\end{equation}}

\def\bga#1\ega{\begin{gather}#1\end{gather}} 
\def\bgas#1\egas{\begin{gather*}#1\end{gather*}}

\def\bal#1\eal{\begin{align}#1\end{align}} 
\def\bals#1\eals{\begin{align*}#1\end{align*}}

\def \tr{\mbox{tr\hskip 1pt}}
\renewcommand{\vec}[1]{\boldsymbol{\mathbf{#1}}}

\begin{document}

\title{A robust anisotropic hyperelastic formulation for the modelling of soft tissue}
\author{David R.~Nolan$^a$, Artur L.~Gower$^b$, \\ Michel Destrade$^b$, Ray W.~Ogden$^c$, Pat McGarry$^a$\\[12pt]
$^a$Biomedical Engineering, \\
National University of Ireland, Galway, Galway, Ireland\\[6pt]
$^b$School of Mathematics, Statistics and Applied Mathematics,\\
 National University of Ireland, Galway, Galway, Ireland\\[6pt]
 School of Mathematics and Statistics, \\
 University of Glasgow, Glasgow, Scotland}
          
\date{}
\maketitle

\begin{abstract}

The Holzapfel--Gasser--Ogden (HGO) model for anisotropic hyperelastic behaviour of collagen fibre reinforced materials was initially developed to describe the elastic properties of arterial tissue, but is now used extensively for modelling a variety of soft biological tissues. Such materials can be regarded as incompressible, and when the incompressibility condition is adopted the strain energy $\Psi$ of the HGO model is a function of one isotropic and two anisotropic deformation invariants.
A compressible form (HGO-C model) is widely used in finite element simulations whereby the isotropic part of $\Psi$ is decoupled into volumetric and isochoric parts and the anisotropic part of $\Psi$ is expressed in terms of isochoric invariants. Here, by using three simple deformations (pure dilatation, pure shear and uniaxial stretch), we demonstrate that the compressible HGO-C formulation does not correctly model compressible anisotropic material behaviour, because the anisotropic component of the model is insensitive to volumetric deformation due to the use of isochoric anisotropic invariants.   
In order to correctly model compressible anisotropic behaviour we present a modified anisotropic (MA) model, whereby the full anisotropic invariants are used, so that a volumetric anisotropic contribution is represented. The MA model correctly predicts an anisotropic response to hydrostatic tensile loading, whereby a sphere deforms into an ellipsoid. It also computes the correct anisotropic stress state for pure shear and uniaxial deformation. 
To look at more practical applications, we developed a finite element user-defined material subroutine for the simulation of stent deployment in a slightly compressible artery. Significantly higher stress triaxiality and arterial compliance are computed when the full anisotropic invariants are used (MA model) instead of the isochoric form (HGO-C model).
\end{abstract}


\section*{Nomenclature}

\noindent $\vec{I}$ -- identity tensor
\\
 $\Psi $ -- Helmholtz free-energy (strain-energy) function
\\
 $\Psi_{\mathrm{vol}}$ -- volumetric contribution to the free energy
 \\
 $\Psi_\mathrm{aniso}$ -- anisotropic contribution to the free energy
\\
 $\overline{\Psi}_\mathrm{iso}$ -- isotropic contribution to the isochoric free energy
\\
 $\overline{\Psi}_\mathrm{aniso}$ -- anisotropic contribution to the isochoric free energy
\\
 $\vec{\sigma}$ -- Cauchy stress
\\
 $\vec{\sigma'}$ -- deviatoric Cauchy stress
\\
 $q$ -- von Mises equivalent stress
\\
 $\sigma_{hyd}$ -- hydrostatic (pressure) stress
\\
 $\vec{F}$ -- deformation gradient
\\
 $J$ -- determinant of the deformation gradient; local ratio of volume change
\\
 $\vec{C}$ -- right Cauchy--Green tensor
\\
 $I_{1}$ -- first invariant of $\vec{C}$
\\
 $I_{4,6}$ -- anisotropic invariants describing the deformation of reinforcing fibres
\\
 $\overline{\vec{F}}$ -- isochoric portion of the deformation gradient
\\
 $\overline{\vec{C}}$ -- isochoric portion of the right Cauchy--Green deformation tensor
\\
 $\overline{I}_{1}$ -- first invariant of $\overline{\vec{C}}$
\\
 $\overline{I}_{4,6}$ -- isochoric anisotropic invariants
\\
 $\vec{a}_{0i},\,i=4,6$ -- unit vector aligned with a reinforcing fibre in the reference configuration
\\
 $\vec{a}_{i},\,i=4,6$ -- updated (deformed) fibre direction ($=\vec{F}\vec a_{0i}$)
\\
 $ \kappa_{0}$ -- isotropic bulk modulus
\\
 $ \mu_{0}$ -- isotropic shear modulus
\\
 $ k_{i},\,i=1,2$ -- anisotropic material constants
\\
 $ \nu $ -- isotropic Poisson's ratio
\\
\\
Bold uppercase symbols represent second order tensors, bold lowercase symbols represent vectors and un-bold symbols represent scalars.


\section{ Introduction}
\label{intro}


The anisotropic hyperelastic constitutive model proposed by \cite{Holzapfel2000} (henceforth referred to as the HGO model) is used extensively to model collagen fibre-reinforced biological materials, even more so  now that it has been implemented in several commercial and open-source Finite Element (FE) codes for the simulation of soft tissue elasticity. 

The constitutive equation builds upon previously published transversely isotropic constitutive models (e.g. \cite{Weiss1996}) and reflects the structural components of a typical biological soft tissue, and hence its strain-energy density consists of two mechanically equivalent terms accounting for the anisotropic contributions of the reinforcing fibre families, in addition to a term representing the isotropic contribution of the ground matrix in which the fibres are embedded.  Also, it assumes that the collagen fibres do not support compression, and hence they provide a mechanical contribution only when in tension (this may be taken care of by pre-multiplying each anisotropic term with a Heaviside, or ``switching'', function).

For the original incompressible HGO model the strain energy $\Psi$ is expressed as a function of one isochoric isotropic deformation invariant (denoted $\overline I_1$) and two isochoric anisotropic invariants (denoted $\overline I_4$ and $\overline I_6$). A Lagrange multiplier is used to enforce incompressibility \citep*{Holzapfel2000}.
Once again it should be stressed that the original HGO model is intended only for the simulation of incompressible materials.

A modification of the original HGO model commonly implemented in finite element codes entails the replacement of the Lagrange multiplier penalty term with an isotropic hydrostatic stress term that depends on a user specified bulk modulus. This modification allows for the relaxation of the incompressibility condition and we therefore refer to this modified formulation as the HGO-C (compressible) model for the remainder of this study.

The HGO-C model has been widely used for the finite element simulation of many anisotropic soft tissues. For example, varying degrees of compressibility have been reported for cartilage in the literature (e.g. \cite{Guilak1995, Smith2001}). It has been modelled as a compressible material using the HGO-C model (e.g. \cite{Pena2007} used a Poisson's ratio, $\nu = 0.1$  and \cite{Perez2006} used $\nu=0.1$ and $\nu=0.4$). To date, material compressibility of arterial tissue has not been firmly established. Incompressibility was assumed by the authors of the original HGO model and in subsequent studies (e.g. \cite{Kiousis2009}). However many studies model arteries as compressible or slightly compressible (e.g. \cite{Cardoso2014} $\nu=0.33-0.43$ and \cite{Iannaccone2014} $\nu=0.475$). In addition to arterial tissue the nucleus pulposus of an inter-vertebral disc has been modelled as a compressible anisotropic material using the HGO-C model (e.g. \citep{Maquer2014} $\nu=0.475$). Furthermore the HGO-C formulation has been used to simulate growth of anisotropic biological materials, where volume change is an intrinsic part of a bio-mechanical process (e.g. \citep{Huang2012} $\nu=0.3$). 
However, the enforcement of perfect incompressibility may not be readily achieved in numerical models. As an example, the finite element solver Abaqus/Explicit assigns a default Poisson's ratio of 0.475 to "incompressible" materials in order to achieve a stable solution \citep{Abaqus2010} and in this case the HGO-C model must be used (e.g. \cite{Conway2012, Famaey2012}).
Despite the widespread use of the HGO-C model, its ability to correctly simulate anisotropic compressible material behaviour has not previously been established.

%
%

\begin{itemize}
\item The first objective of this study is to demonstrate that the HGO-C formulation does not correctly model an anisotropic compressible hyperelastic material. 
\end{itemize}
Recently, \cite{Vergori2013} showed that under hydrostatic tension, a sphere consisting of a slightly compressible HGO-C material expands into a larger sphere instead of deforming into an ellipsoid.  It was suggested that this effectively isotropic response is due to the isochoric anisotropic invariants $\overline{I}_i$ being used in the switching function instead of the full invariants $I_i,\,i=4,6$.  However, in the current paper we show that the problem emerges fundamentally because there is no dilatational contribution to the anisotropic terms of $\Psi$. In fact, modifying only the ``switching criterion'' for fibre lengthening is not sufficient to fully redress the problem.
\begin{itemize}
\item The second objective of the study is to implement a modification of the HGO-C model so that correct anisotropic behaviour of compressible materials is achieved.
\end{itemize}

This modified anisotropic (MA) model uses the full form of the anisotropic invariants and through a range of case studies we show this leads to the correct computation of stress in contrast to the widely used HGO-C model.

%
%
The paper is structured as follows.  In Section \ref{section2} we demonstrate and highlight the underlying cause of the insensitivity of the anisotropic component of the  HGO-C model to volumetric deformation in compressible materials.
We demonstrate that the modification of the model to include the full form of the anisotropic invariants corrects this deficiency.  In Section \ref{section3} we show how the HGO-C model yields unexpected and unphysical results for pure in-plane shear and likewise in \ref{section4} for simple uniaxial stretching, in contrast to the modified model.  We devote Section \ref{section5} to two Finite Element biomechanics case studies, namely pressure expansion of an artery and stent deployment in an artery, and illustrate the significant differences in computed results for the HGO-C model and the modified model.
Finally, we provide some concluding remarks and discussion points in Section \ref{conclusion}.


\section{Theory: Compressible Anisotropic Hyperelastic Constitutive Models }
\label{section2}



\subsection{ HGO-C Model for Compressible Materials }

The original HGO model is intended for incompressible materials. However a variation of the HGO model whereby a bulk modulus is used instead of a penalty term has been implemented in a number of FE codes. Several authors have used this formulation t model compressible anisotropic materials bu using a relatively low value of bulk modulus. An important objective of this paper is to highlight that this HGO-C formulation does not correctly model compressible anisotropic material behaviour.

The kinematics of deformation are described locally in terms of the deformation gradient tensor, denoted $\vec{F}$, relative to some reference configuration.  The right Cauchy--Green tensor is defined by $\vec{C}=\vec{F}^\mathrm{T}\vec{F}$, where $^\mathrm{T}$ indicates the transpose of a second-order tensor. 

Hyperelastic constitutive models used for rubber-like materials often split the local deformation into volume-changing (volumetric) and volume-preserving (isochoric, or deviatoric) parts. 
Accordingly the deformation gradient ${\vec F}$ is decomposed multiplicatively as follows: 

\begin{equation}
\label{GrindEQ__1_} 
\vec{F}=\left(J^{\frac{1}{3}}{\vec I}\right)\overline{\vec{F}},
\end{equation}
where $J$ is the determinant of ${\vec F}$. 
The term in the brackets represents the volumetric portion of the deformation gradient and $\overline{\vec{F}}$ is its isochoric portion, such that $\det  (\overline{\vec F})=1$ at all times.

Suppose that the material consists of an isotropic matrix material within which are embedded two families of fibres  characterized by two preferred directions in the reference configuration defined in terms of two unit vectors $\vec{a}_{0i},\,i=4,6$.  With $\vec{C}$, $J$ and $\vec{a}_{0i}$ are defined the invariants
\begin{equation}
I_1=\tr (\vec{C}),\quad I_4=\vec{a}_{04}\cdot(\vec{C}\vec{a}_{04}),\quad I_6=\vec{a}_{06}\cdot(\vec{C}\vec{a}_{06}),
\label{eq_invariants}
\end{equation}
\begin{equation}
\overline{I}_1= J^{-2/3} I_{1},\quad \overline{I}_4= J^{-2/3} I_{4},\quad \overline{I}_6= J^{-2/3} I_{6}
\label{eq_isochoric_invariants},
\end{equation}
where $\overline{I}_i$ ($i=1,4,6$) are the isochoric counterparts of $I_i$. The HGO model proposed by \cite{Holzapfel2000} for collagen reinforced soft tissues additively splits the strain energy $\Psi$ into volumetric, isochoric isotropic and isochoric anisotropic terms,
\begin{equation}
\label{GrindEQ__3_}
\Psi \left(\vec C, \vec a_{04}, \vec a_{06} \right) = \Psi_\text{vol} \left(J\right) + \overline{\Psi}_\text{iso} \left(\overline{\vec C}\right) + \overline{\Psi}_\text{aniso} \left(\overline{\vec C}, \vec a_{04}, \vec a_{06} \right), 
\end{equation}
where ${\overline{\Psi }}_\text{iso}$ and ${\overline{\Psi }}_\text{aniso}$  are the isochoric isotropic and isochoric anisotropic free-energy contributions, respectively, and $ \overline{\vec C} = J^{-2/3}\vec C$ is the isochoric right Cauchy--Green deformation tensor.
 
In numerical implementations of the model \citep{Abaqus2010, Adina2005, Gasser2002}, the volumetric and isochoric isotropic terms are represented by the slightly compressible neo-Hookean hyperelastic free energy 
\begin{equation}
\label{neoH}
\Psi_\text{vol}(J) = \dfrac{1}{2} \kappa_{0} \left(J-1\right)^2, \qquad
\overline \Psi_\text{iso}(\overline{\vec C}) = \dfrac{1}{2}\mu_{0}(\overline I_1 - 3),
\end{equation}
where $\kappa_{0}$ and $\mu_{0}$ are the bulk and shear moduli, respectively, of the soft isotropic matrix. Of course one may write \eqref{neoH} in terms of the full invariants also, using the results from \eqref{eq_isochoric_invariants}.

The isochoric anisotropic free-energy term is prescribed as
\begin{equation}
 \overline{\Psi}_\text{aniso} \left(\overline{\vec C}, \vec a_{04}, \vec a_{06} \right)
 = \dfrac{k_1}{2k_2} \sum_{i=4,6} \{ \exp  [ k_2 \left(\overline{I}_i - 1\right)^2 ] - 1\},
 \label{GrindEQ__4_}
\end{equation}
where $k_1$ and $k_2$ are positive material constants which can be determined from experiments.

For a general hyperelastic material with free energy $\Psi$  the Cauchy stress is given by
\begin{equation}
 \vec{\sigma} = \dfrac{1}{J}\vec{F} \dfrac{\partial \Psi }{\partial \vec{F}}.
 \label{GrindEQ__dphidF_}
\end{equation}
For the Cauchy stress derived from $\Psi$ above, we have the decomposition $\vec{\sigma} = \vec{\sigma}_\text{vol} + \overline{\vec{\sigma}}_\text{iso} + \overline{\vec \sigma}_\text{aniso}$, where
\begin{equation} \label{devia}
\vec{\sigma}_\text{vol} = \kappa_{0} (J-1)\vec{I}, \quad
\overline{\vec{\sigma}}_\text{iso} = \mu_{0} J^{-1}\left(\overline{\vec{B}}  - \tfrac{1}{3}\overline{I}_1\vec{I}\right),
\end{equation}
with $\overline{\vec{B}} = \overline{\vec{F}} \, \overline{\vec{F}}^\mathrm{T}$, and
\begin{equation}
\overline{\vec{\sigma}}_\text{aniso} = 2k_1J^{-1} \sum_{i=4,6} \left(\overline I_i - 1\right) \exp  [k_2 \left(\overline I_i - 1\right)^2]  (\overline{\vec{a}}_i \otimes \overline{\vec{a}}_i - \tfrac{1}{3} \overline I_i \vec{I} ),
  \label{GrindEQ__strhgo_}
\end{equation}
where $\overline{\vec{a}}_i = \overline{\vec{F}} \vec a_{0i}$.
This slightly compressible implementation is referred to as the HGO-C model henceforth. 

The original incompressible HGO model by \cite{Holzapfel2000} specified that for arteries the constitutive formulation should be implemented for incompressible materials. In that limit, $\kappa_{0} \to \infty$, $(J-1) \to 0$ while the product of these two quantities becomes an indeterminate Lagrange multiplier, $p$, and the volumetric stress assumes the form, $\vec{\sigma}_{\text{vol}}=-p \vec{I}$. Indeed the original incompressible HGO model can equally be expressed in terms of the full invariants $I_4$ and $I_6$ (with $J \to 1$) (e.g., \cite{Holzapfel2004}).

However, in the case of the HGO-C implementation, if $ \kappa_{0} $ is not fixed numerically at a large enough value, then slight compressibility is introduced into the model.
The key point of this paper is that the isochoric anisotropic term $\overline \Psi_\text{aniso}$ defined in \eqref{GrindEQ__4_} does not provide a full representation of the anisotropic contributions to the stress tensor for slightly compressible materials. In Section \ref{section2-MA} we introduce a simple modification of the anisotropic term to account for material compressibility.

%

\subsection{Pure dilatational deformation}
\label{section2-PD}


First we consider the case of the HGO-C material subjected to a \emph{pure dilatation} with stretch $\lambda=J^{1/3}$, so that
\begin{equation}
\vec{F} = \lambda \vec{I}, \quad \vec{C} = \lambda^2 \vec{I}, \quad  J = \lambda^3.
\label{GrindEQ__9_}
\end{equation}
We expect that an anisotropic material requires an anisotropic stress state to maintain the pure dilatation. However, calculation of the invariants $I_i$ and ${\overline{I}}_i$ yields
\begin{equation}
 I_i = \vec{a}_{0i} \cdot( \vec{C} \vec{a}_{0i}) = \lambda^2, \quad  \overline{I}_i = J^{-2/3} I_i = 1, \quad i=4,6,
\end{equation}
so that while $I_i$ is indeed the square of the fibre stretch and changes with the magnitude of the dilatation, its isochoric counterpart $\overline{I}_i$ is always unity.
Referring to \eqref{GrindEQ__strhgo_}, it is clear that the entire anisotropic contribution to the stress \eqref{GrindEQ__dphidF_} disappears (i.e. $\overline{\vec{\sigma}}_\text{aniso} \equiv \vec{0}$), and the remaining active terms are the isotropic ones. Thus, \emph{under pure dilatation, the HGO-C model computes an entirely isotropic state of stress.}


\subsection{Applied hydrostatic stress}
\label{section2-AHS}


Now we investigate the reverse question: what is the response of the HGO-C material to a \emph{hydrostatic stress},
\begin{equation}
\vec{\sigma} = \sigma \vec{I},
\end{equation}
where $\sigma>0$ under tension and $\sigma<0$ under pressure?
In an anisotropic material, we expect the eigenvalues of $\vec{C}$, the squared principal stretches, $\lambda_1^2$, $\lambda_2^2$, $\lambda_3^2$ say, to be distinct.
Hence, if the material is slightly compressible, then a sphere should deform into an ellipsoid \citep{Vergori2013} and a cube should deform into a hexahedron with non-parallel faces \citep{NiAnnaidh2013b}.

However, in the HGO-C model the $\overline{\Psi}_\text{aniso}$ contribution is switched on only when $\overline{I}_i$ (not $I_i$) is greater than unity.
 \cite{Vergori2013} showed that in fact $\overline{I}_i$ is always less than or equal to one in compression and in expansion under hydrostatic stress, so that the HGO-C response is isotropic, contrary to physical expectations. Then we may ask if removal of the switching function circumvents this problem so that anisotropic response is obtained.

With the fibres taken to be mechanically equivalent and aligned with $\vec{a}_{04}= (\cos \Theta,  \sin\Theta,0)$ and  $\vec{a}_{06}=(\cos \Theta, - \sin\Theta,0)$ in the reference configuration, we have, by symmetry, $I_6=I_4$ and $\overline{I}_6=\overline{I}_4$ and $\overline{\Psi}_6=\overline{\Psi}_4$, where the subscripts $4$ and $6$ on $\overline{\Psi}$ signify partial differentiation with respect to $\overline{I}_4$ and $\overline{I}_6$, respectively.  Similarly, in the following the subscript $1$ indicates differentiation with respect to $\overline{I}_1$. For this special case,
\cite{Vergori2013} showed that the stretches arising from the application of a hydrostatic stress are
 \begin{align} \label{lambdas}
& \lambda_1=J^{1/3}\left[\frac{\overline \Psi_1(\overline \Psi_1+2\overline \Psi_4\sin^2\Theta)}{(\overline \Psi_1+2\overline \Psi_4\cos^2\Theta)^2}\right]^\frac{1}{6},\notag \\[8pt]
&  \lambda_2=J^{1/3}\left[\frac{\overline \Psi_1(\overline \Psi_1+2\overline \Psi_4\cos^2\Theta)}{(\overline \Psi_1+2\overline \Psi_4\sin^2\Theta)^2}\right]^\frac{1}{6},\notag \\[8pt]
& \lambda_3=J^{1/3}\left[\frac{\overline \Psi_1^2+2\overline \Psi_1\overline \Psi_4+\overline \Psi_4^2\sin^22\Theta}{\overline \Psi_1^2}\right]^\frac{1}{6}.
 \end{align}
Explicitly,
\begin{equation}
\overline{\Psi}_1 = \dfrac{\partial \overline{\Psi}}{\partial \overline{I}_1} = \frac{1}{2}\mu_{0}, \quad
\overline{\Psi}_4 =  \dfrac{\partial \overline{\Psi}}{\partial \overline{I}_4} = k_1 (\overline{I}_4 -1) \exp[   k_2 \left(\overline{I}_4 - 1\right)^2].
\end{equation}
Looking at \eqref{lambdas}, we see that there is a solution to the hydrostatic stress problem where the stretches are unequal, so that a sphere deforms into an ellipsoid.
However, there is also another solution: that for which $\overline{I}_4 \equiv 1$, in which case, $\overline{\Psi}_4 \equiv 0$ by the above equation, and then $\lambda_1=\lambda_2=\lambda_3= J^{1/3}$ by \eqref{lambdas}.  Thus, a sphere then deforms into another sphere.

Of those (at least) two possible paths, FE solvers converge upon the isotropic solution.
One possible explanation for this may be that the initial computational steps calculate strains in the small-strain regime.      
In that regime, \cite{Vergori2013} showed that all materials with a decoupled volumetric/isochoric free-energy behave in an isotropic manner when subject to a hydrostatic stress. 
Hence the first computational step brings the deformation on the isotropic path, and $\overline I_4 =1$ then, and subsequently.
In Section \ref{section2-PD} and Section \ref{section2-AHS} we have thus demonstrated that the use of an isochoric form of the anisotropic strain energy $ \overline{\Psi}_\text{aniso} $ from the HGO model in the HGO-C model cannot yield a correct response to pure dilatation or applied hydrostatic stress.


\subsection{ Modified Anisotropic Model for Compressible Materials}
\label{section2-MA}

In order to achieve correct anisotropic behaviour for compressible materials we introduce a modification to the anisotropic term of the HGO model, whereby the anisotropic strain energy is a function of the `total' right Cauchy--Green deformation tensor $\vec{C}$, rather than its isochoric part $\overline{\vec{C}}$, so that

\begin{equation}
 \Psi \left(J,{\vec{C}}, \vec{a}_{04}, \vec{a}_{06}\right)= \Psi_\text{vol} \left(J\right) +  \Psi_\text{iso} (J,\vec{C}) + \Psi_\text{aniso} (\vec{C},\vec{a}_{04}, \vec{a}_{06} ),
 \label{GrindEQ__14_}
\end{equation}

where the expressions for strain energy density terms $ \Psi_{\text{vol}} $ and $\overline \Psi_\text{iso}$ are the same as those in \eqref{neoH}, and
\begin{equation}
 \Psi_\text{aniso}\left(\vec{C}, \vec{a}_{04}, \vec{a}_{06}\right) = \dfrac{k_1}{2k_2} \sum_{i = 4,6} \{\exp  [k_2 \left(I_i - 1\right)^2 ] -1\}.
 \label{GrindEQ__15_}
\end{equation}
This modification to the HGO-C model is referred to as the \emph{modified anisotropic} (MA) \emph{model} hereafter.
Combining \eqref{neoH}, \eqref{GrindEQ__14_} and \eqref{GrindEQ__15_}, the Cauchy stress for the MA model is determined using \eqref{GrindEQ__dphidF_} and the decomposition $\vec{\sigma} = \vec{\sigma}_\text{vol} + \overline{\vec{\sigma}}_\text{iso} + {\vec \sigma}_\text{aniso}$ resulting in the expression:

\begin{equation}
\vec{\sigma}= \kappa_{0} (J-1)\vec{I} + \mu_{0} J^{-5/3} \left(\vec B  - \tfrac{1}{3} I_1\vec I\right) + 2k_1\sum_{i=4,6} \left(I_i - 1\right) \exp  [k_2{\left(I_i-1\right)}^2] \vec{a}_i \otimes \vec{a}_i .
\label{GrindEQ__stressmp_}
\end{equation}

where $\vec{a}_i = \vec{F} \vec{a}_{0i},\,i=4,6$. Now it is easy to check that in the cases of a pure dilatation and of a hydrostatic stress, the MA model behaves in an anisotropic manner, because the term $I_i-1\neq 0$ and hence $\Psi_\text{aniso}\neq 0$ and $\vec{\sigma}_\text{aniso} \neq\vec{0}$.  This resolves the issues identified above for the HGO-C model.

We have developed a user-defined material model (UMAT) Fortran subroutine to implement the MA formulation for the Abaqus/Standard FE software.
The FE implicit solver requires that both the Cauchy stress and the consistent tangent matrix (material Jacobian) are returned by the subroutine.
Appendix A gives the details of the consistent tangent matrix.

\begin{figure}
\centering
 \includegraphics[width=\textwidth ]{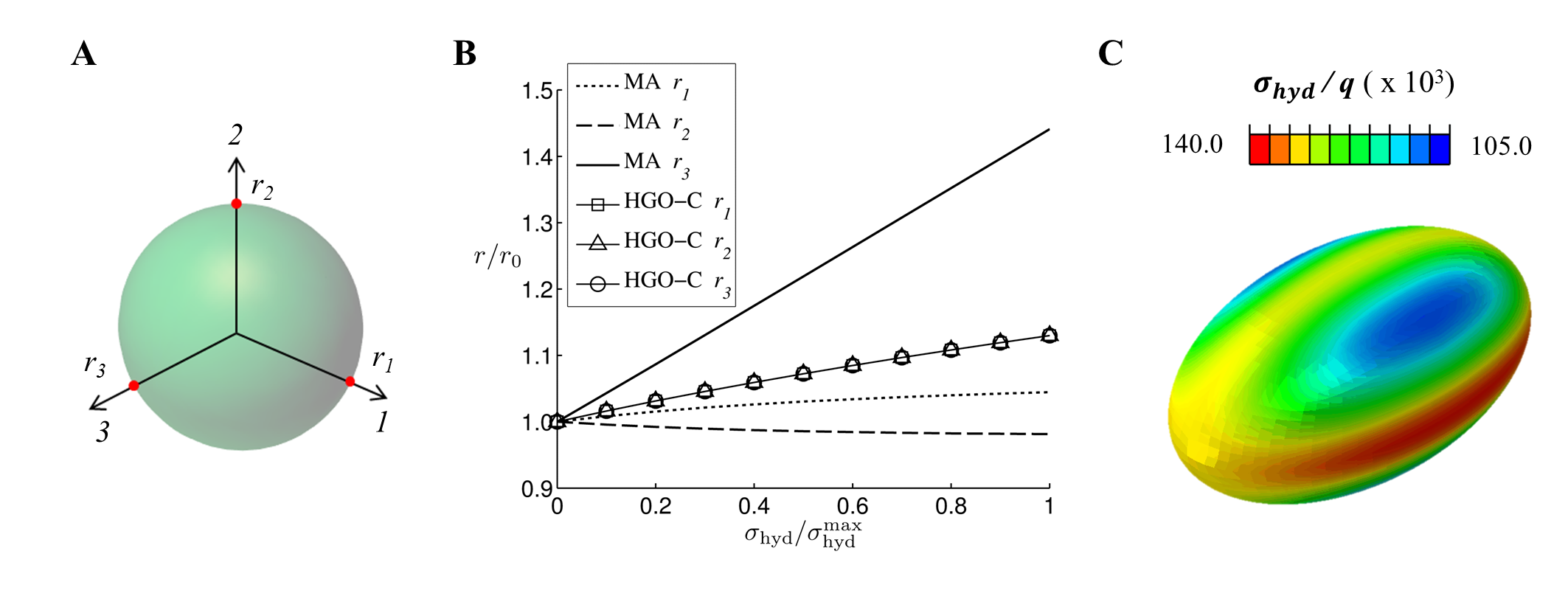}
\caption{ \textbf{A)} Schematic of an undeformed sphere highlighting three radii on orthogonal axes, 1-2-3, centred at the sphere origin. Two families of fibres are contained in the $(1,2)$ plane and symmetric about the 1-axis. \textbf{B)} Computed (deformed/undeformed) ratios $(r/r_0)$ of the orthogonal radii for both MA and HGO-C models versus the ratio $\sigma_{\text{hyd}}/\sigma_{\text{hyd}}^{\mathrm{max}}$. Note that the deformation computed for the HGO-C model incorrectly remains spherical. \textbf{C)} Deformed ellipsoidal shape computed for the MA model; contours illustrate the inhomogeneous distribution of stress triaxiality ($\sigma_{\text{hyd}}/q$) throughout the deformed body. }
\label{fig:sphere}
\end{figure}

We have used the above subroutine to repeat the simulations of expansion of a sphere under hydrostatic tension of \cite{Vergori2013}, this time using the MA formulation.
Again two families of fibres are assumed, lying in the $(1,2)$ plane and symmetric about the \textit{1}-axis (the sphere and axes are shown in Figure \ref{fig:sphere}A).
The displacements of points on the surface of the sphere at the ends of three mutually orthogonal radii with increasing applied hydrostatic tension are shown in Figure \ref{fig:sphere}B.
Clearly the sphere deforms into an ellipsoid with a major axis oriented in the 3-direction and a minor axis oriented in the 1-direction, confirming the simulation of orthotropic material behaviour.
The distribution of stress triaxiality in the deformed ellipsoid, measured by ${\sigma_{\text{hyd}}}/{q}$, is shown in Figure \ref{fig:sphere}C, where $\sigma_{\text{hyd}} \equiv \tr (\vec{\sigma})/3$ is the hydrostatic stress and $q \equiv \sqrt{3/2\,\vec{ \sigma}' :\vec{\sigma}'}$ is the von Mises equivalent stress, $\vec{\sigma}'$ being the deviatoric Cauchy stress tensor. Clearly an inhomogeneous stress state is computed in the deformed body.

The results shown in Figure \ref{fig:sphere} contrast sharply with the equivalent simulations using the HGO-C model \citep{Vergori2013} superimposed in Figure \ref{fig:sphere}B for comparison.
In that case a similar fibre-reinforced sphere is shown to deform into a larger sphere with a homogeneous stress distribution, indicative of isotropic material behaviour.

\section{ Analysis of Pure Shear}
\label{section3}


A pure dilatation and a hydrostatic stress each represent a highly idealized situation, unlikely to occur by themselves in soft tissue \textit{in vivo}.
This section highlights the unphysical behaviour can also emerge for common modes of deformation if the anisotropic terms are based exclusively on the isochoric invariants.
Considering once again the general case of a compressible anisotropic material, we analyse the response of the HGO-C and MA models to pure in-plane shear. Regarding the out-of-plane boundary conditions, we first consider the case of plane strain (Section 3.1). Even though this deformation is entirely isochoric the HGO-C model yields incorrect results. We then consider the case of plane stress (Section 3.2), and again demonstrate that the HGO-C model yields incorrect results. 
By contrast, we show that the MA model computes a correct stress state for all levels of compressibility and specified deformations. In the following calculations we assume a shear modulus, $ \mu_{0} = 0.05$\,MPa and anisotropic material constants $k_1 = 1$\,MPa and $k_2 = 100$.


\subsection{Plane strain pure shear}
\label{_pure_shear_pe_}

\begin{figure}[!t]
{\centering
\includegraphics*[width=\linewidth, keepaspectratio=false]{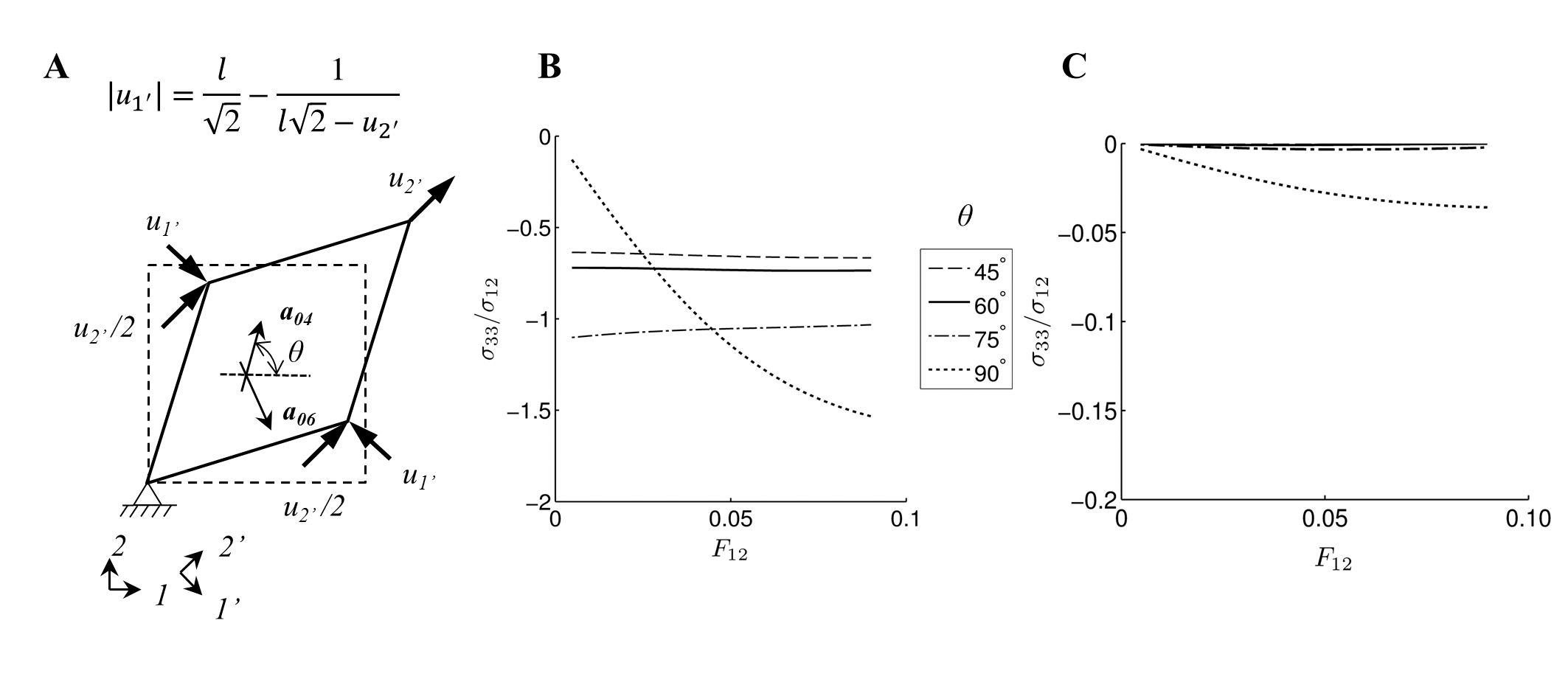}
}
\caption{
\textbf{A)} Schematic illustrating the kinematics of the pure shear deformation of the $(1,2)$ section of a unit cube.
Note the rotated coordinate system $(1',2')$, orientated at $45^\circ$ to the $(1,2)$ axes, used to specify the vertex displacement components $u_{1'}$ and $u_{2'}$.
Note also the vectors $\vec{a}_{0i}, \,i=4,6$, indicating the directions of the two families of  fibres, with angle $\theta$.
Results are displayed for a range of fibre orientations with $\theta$ from $\pm 45^\circ$ to $\pm 90^\circ$ with respect to the $(1,2)$ coordinate system.
\textbf{B)} Computed stress ratio  ${\sigma }_{33}/{\sigma }_{12}$ versus $F_{12}$ for the HGO-C model,  illustrating significant negative (compressive) stresses in the out-of-plane direction. \textbf{C)} Computed stress ratio versus $F_{12}$ for the MA model, illustrating very small negative (compressive) stresses in the out-of-plane direction (an order of magnitude lower than for the HGO-C model). }
\label{fig_shear01}
\end{figure}

With restriction to the $(1,2)$ plane we now consider the plane strain deformation known as \emph{pure shear}, maintained by the application of a suitable Cauchy stress.
In particular,  we take the deformation gradient for this deformation to have components
\begin{equation}
\mathbf{F}=\left[\begin{array}{ccc}
\sqrt{F^2_{12}+1} & F_{12} & 0 \\
F_{12} & \sqrt{F^2_{12}+1} & 0 \\
0 & 0 & 1 \end{array}\right],
\label{GrindEQ__defgrd}
\end{equation}
where $F_{12}$ is a measure of the strain magnitude.  Figure \ref{fig_shear01}A depicts the deformation of the $(1,2)$ square cross section of a unit cube, which deforms into a parallelogram symmetric about a diagonal of the square.  The deformation corresponds to a stretch $\lambda=\sqrt{F_{12}^2+1} +F_{12}$ along the leading diagonal with a transverse stretch $\lambda^{-1}=\sqrt{F_{12}^2+1} -F_{12}$.  We can think of the deformation arising from displacement components applied to the vertices of the square, as indicated in Figure \ref{fig_shear01}A.
Two families of fibres, with reference unit vectors $\vec{a}_{04}$ and $\vec{a}_{06}$ are assumed to lie in the $(1,2)$ plane, as illustrated in Figure \ref{fig_shear01}A, oriented with angles $\pm\theta$ to the $1$ axis. We perform some calculations for a range of fibre orientations for each of the HGO-C and MA models.

First we note that although, for this specific case, the free energies of the HGO-C and the MA models coincide (because $J= 1$ and hence $I_4=\overline{I}_4$), the corresponding stress tensors are very different.  This is due to the ``deviatoric'' form of the anisotropic stress contribution that emerges for the HGO-C model, as in the final term of \eqref{GrindEQ__strhgo_}, compared with the final term of  \eqref{GrindEQ__stressmp_}.
It gives rise to a significant negative (compressive) out-of-plane stress component $\sigma_{33}$ which is comparable in magnitude to $\sigma_{12}$, as shown in Figure \ref{fig_shear01}B. Such a negative stress is anomalous in the sense that for large $\kappa_{0}$ the result for the incompressible limit should be recovered, but it is not. 
Indeed, if we start with the incompressible model we obtain $\sigma_{33} = \mu_{0} - p$, which is independent of $\sigma_{12}$. However, as \eqref{GrindEQ__defgrd} represents a kinematically prescribed isochoric deformation, the volumetric stress in the HGO-C model goes to zero and does not act as the required Lagrange multiplier.

By contrast, the out-of-plane compressive normal stress component $\sigma_{33}$ computed for the MA model is at least an order of magnitude lower than the in-plane shear stress component $\sigma_{12}$ (Figure \ref{fig_shear01}C), and is close to zero for most fibre orientations.   This is consistent with the incompressible case because, since $p$ is arbitrary it may be chosen to be $\mu_{0}$ so that $\sigma_{33}=0$. This is what might be expected physically, given that the fibres and the deformations are confined to the $(1,2)$ plane.

\begin{figure}[!t]
{\centering
\includegraphics*[width=\linewidth, keepaspectratio=false]{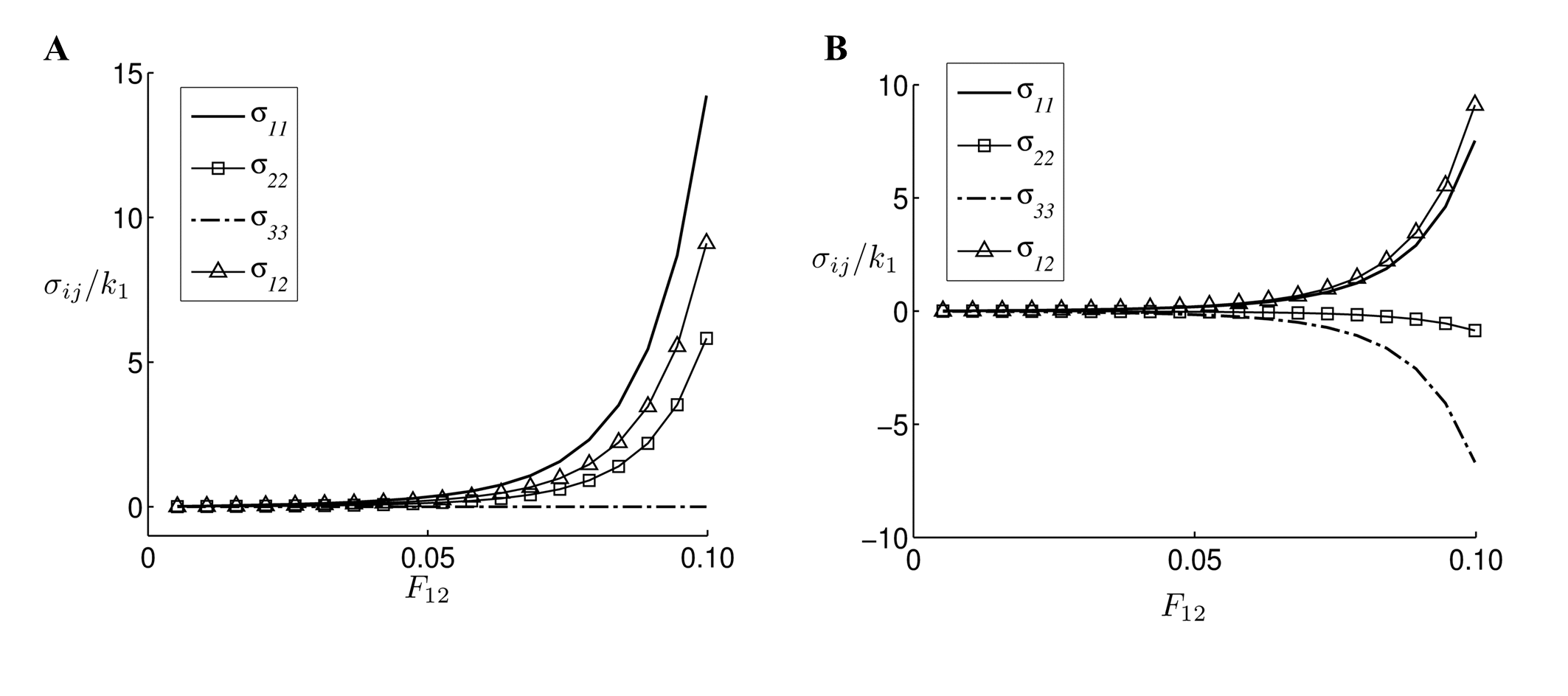}
}
\caption{Dimensionless stress components $\sigma_{ij}/k_1$ versus $F_{12}$ for the case of a single family of fibres orientated at $\theta=30^\circ$. \textbf{A)} MA model; \textbf{B)} HGO-C model.}
\label{fig_shear02}
\end{figure}

Because of the deviatoric component of the stress tensor emerging from the HGO-C model, the trace of the Cauchy stress  is always zero when $J=1$ as equations \eqref{devia} and \eqref{GrindEQ__strhgo_} will confirm.  
By contrast, the trace of the Cauchy stress  is not zero for the MA model.  Hence the in-plane stress components are significantly different from those for the HGO-C model, as shown in Figures \ref{fig_shear02}A and \ref{fig_shear02}B, respectively, for the case of a single fibre family with $\theta=30^\circ$.

\subsection{Plane stress pure shear}


The kinematically prescribed isochoric deformation in Section \ref{_pure_shear_pe_} is volume conserving and makes the $\Psi_{\text{vol}}$ terms equal to zero.
We modify the out-of-plane boundary condition to enforce a plane stress ($\sigma_{33}=0$) simulation
This allows a compressible material to deform of out-of-plane.

A plane stress pure shear deformation is given as
\begin{equation}
\mathbf{F}=\left[\begin{array}{ccc}
\sqrt{F^2_{12}+1} & F_{12} & 0 \\
F_{12} & \sqrt{F^2_{12}+1} & 0 \\
0 & 0 & F_{33} \end{array}\right],
\label{GrindEQ__defgrd-again}
\end{equation}
where the out of plane stretch component $F_{33}$ in general is not equal to $1$, so that the deformation is not in general isochoric.
If the bulk modulus $\kappa_{0}$ is very large compared with the initial shear modulus $\mu_{0}$, then it acts as a Lagrange multiplier to enforce incompressibility, such that $F_{33} = 1$ (at least approximately). If the magnitude of the bulk modulus is reduced, then the material becomes slightly compressible and $F_{33}\neq 1$.
Here we investigate the sensitivity of the stress computed for the HGO-C and MA models to the magnitude of the bulk modulus $\kappa_{0}$.

\begin{figure}[!t]
{\centering
\includegraphics*[width=\linewidth, keepaspectratio=false]{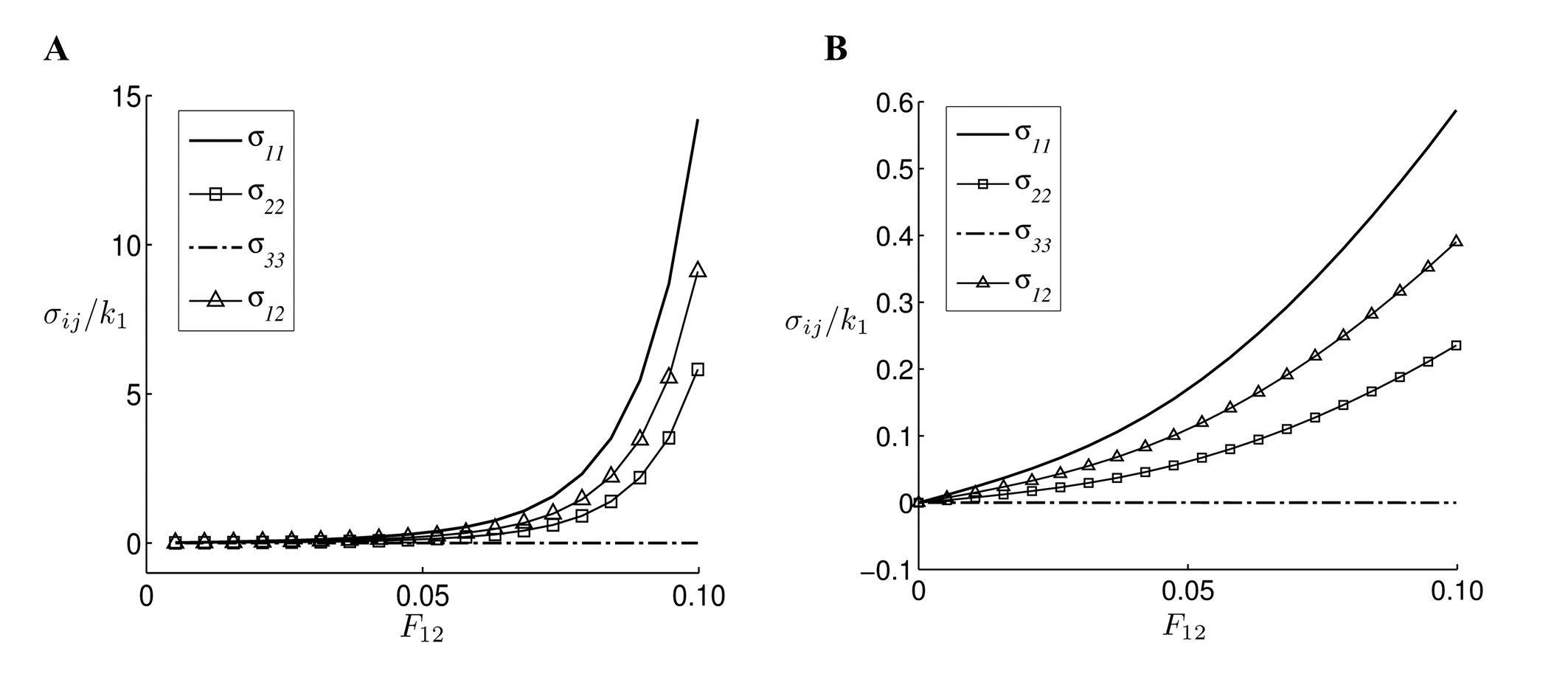}
}
\caption{Dimensionless plots of the normal and in-plane shear Cauchy stress components $\sigma_{ij}/k_1$ versus $F_{12}$ for the case of a single family of fibres orientated at $\theta=30^\circ$. \textbf{A)} Computed stresses for both the HGO-C and MA models with a large bulk modulus $\kappa_{0}/\mu_{0}=2 \times 10^6$ (equivalent to a Poisson ratio of 0.49999975). \textbf{B}) Computed stresses for the HGO-C model with $\kappa_{0}/\mu_{0}=50$ (equivalent to a Poisson ratio of 0.490). Note that the stresses computed for the HGO-C model are an order of magnitude lower in the slightly compressible small bulk modulus case than in the almost incompressible large bulk modulus case.}
\label{fig_shear03}
\end{figure}

First, we consider the almost incompressible case where the ratio of bulk to shear modulus is $\kappa_{0}/\mu_{0}=2 \times 10^{6}$ for the isotropic neo-Hookean component of the model, equivalent to a Poisson ratio of $\nu = 0.49999975$.  The stress components are shown in Figure \ref{fig_shear03}A.  An important point to note is that in this case the deformation is effectively isochoric, because we find $J = F_{33}=1.00006$, and yet the HGO-C model predicts an entirely different stress state from that for the kinematically constrained isochoric deformation of the previous section shown in Figure \ref{fig_shear02}B.  This is because the volumetric term of the free energy now contributes to the trace of the stress tensor, and therefore the high magnitude of bulk modulus effectively acts as a Lagrange multiplier to enforce incompressibility. Indeed for these conditions the HGO-C and MA models behave identically to the original HGO model. However, unlike the HGO-C model, the MA model computes identical stress components for both the kinematically constrained isochoric deformation \eqref{GrindEQ__defgrd} and for the Lagrange multiplier enforced volume preserving deformation \eqref{GrindEQ__defgrd-again}.

If the incompressibility constraint is slightly relaxed, so that $\kappa_{0}/\mu_{0}=50$ ($ \nu = 0.490 $) the HGO-C model computes a very different stress state, as shown in Figure \ref{fig_shear03}B, with stress components being reduced by an order of magnitude.  Thus the HGO-C model is very sensitive to changes in the bulk modulus and, consequently, incompressibility must be enforced by choosing a very large magnitude for  the bulk modulus in order to avoid the computation of erroneous stress states.

By contrast, the MA model computes identical stress states for $\kappa_{0}/\mu_{0}=2 \times 10^6$ and $\kappa_{0}/\mu_{0}=50$ (Figure \ref{fig_shear03}A in both cases).
This response highlights the robustness of the MA model, which computes correct results for all levels of material compressibility (including the incompressible limit).


\section{ Uniaxial stretch}
\label{section4}

\begin{figure}[h]
\centering
 \includegraphics*[width=\linewidth, keepaspectratio=false]{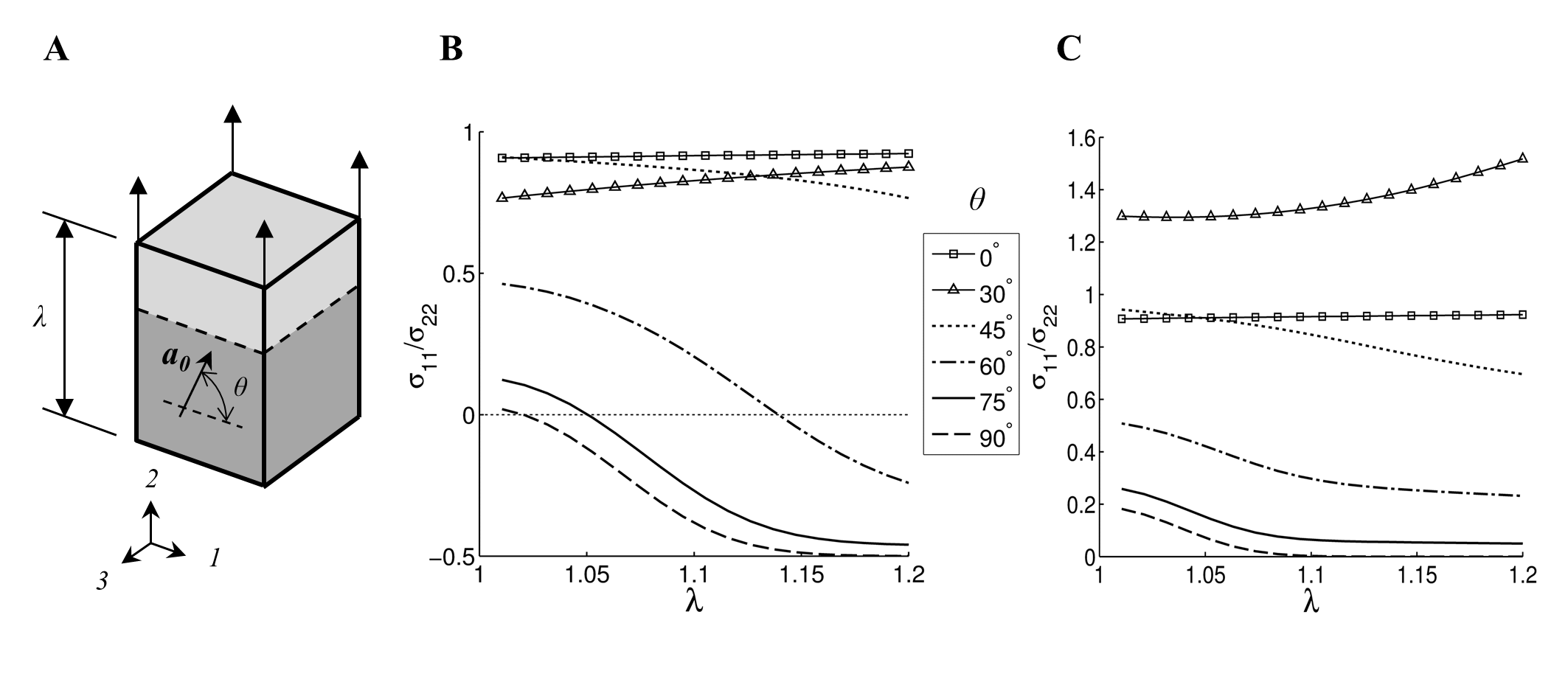}
\caption{
\textbf{A)} Schematic of confined uniaxial stretch ($\lambda_2=\lambda >1, \lambda_1= \lambda_3=1$), showing the fibre family reference directional vector $\vec{a}_0$ in the $(1,2)$ plane.
The ratio of the Cauchy stress components $\sigma_{11}/\sigma_{22}$ is computed  based on a model with a single fibre family and plotted as a function of $\lambda$.   Results are displayed for a range of fibre orientations $\theta$ from $0^\circ$ to $90^\circ$.
\textbf{B)} Computed results for the HGO-C model, illustrating negative (compressive) lateral stresses.
\textbf{C)} Computed results for the MA model, all lateral stresses being positive (tensile).}
\label{fig:uni}
\end{figure}

We now consider a confined uniaxial stretch, as illustrated in Figure \ref{fig:uni}A, where a stretch is imposed in the $2$-direction ($\lambda_2=\lambda>1$) and no lateral deformation is permitted to occur in the $1$- and $3$-directions ($\lambda_1=\lambda_3=1$).
Such a simple deformation may have biomechanical relevance as, for example, in a blood vessel undergoing large circumferential strain, but little or no axial or radial strain.

We derive analytically the stress components for the HGO-C and MA models using the formulas of Section \ref{section2}.
We assume there is a single family of parallel fibres aligned with the reference unit vector $\vec{a}_0$ in the $(1,2)$ plane and with orientation $\theta$ relative to the $1$-axis ranging from $0^\circ$ to $90^\circ$.
We take $\mu_{0} = 0.05$\,MPa, $\kappa_{0}= 1$\,MPa for the slightly compressible neo-Hookean isotropic matrix, and material constants $k_1= 1$\,MPa and $k_2 = 100$ for the fibre parameters.

The ratio of the lateral to axial Cauchy stress components, ${{\sigma }_{11}/\sigma }_{22}$, is plotted as a function of applied stretch $\lambda$ for the HGO-C model (Figure \ref{fig:uni}B) and the MA model (Figure \ref{fig:uni}C).
Results for the HGO-C model exhibit negative (compressive) stresses in the lateral direction for certain fibre orientations. This auxetic effect suggests that the material would expand in the lateral direction in the absence of the lateral constraint and is contrary to expectations, particularly for fibre orientations closer to the axial direction.
In fact, here the computed lateral compressive force is most pronounced when the fibre is aligned in the direction of stretch ($\theta=90^\circ$), where a transversely isotropic response, with exclusively tensile lateral stresses, should be expected.  For all fibres orientated within about $45^\circ$ of the direction of stretch, the lateral stress changes from tensile to compressive as the applied stretch increases.  By contrast to the HGO-C model, the MA model yields exclusively tensile lateral stresses for all fibre orientations (Figure \ref{fig:uni}C).


\section{ Finite Element analysis of realistic arterial deformation}
\label{section5}


Following from the idealized, analytical deformations considered above, we now highlight the practical significance of the errors computed by using the HGO-C model for slightly compressible tissue. We consider, in turn, two Finite Element case studies using \cite{Abaqus2010} to implement the HGO-C and MA models with user-defined material subroutines  (see Appendix A).


\subsection{Pressure expansion of an artery}
\label{pressure}

\begin{figure}[h!]
\centering
 \includegraphics*[width=\linewidth]{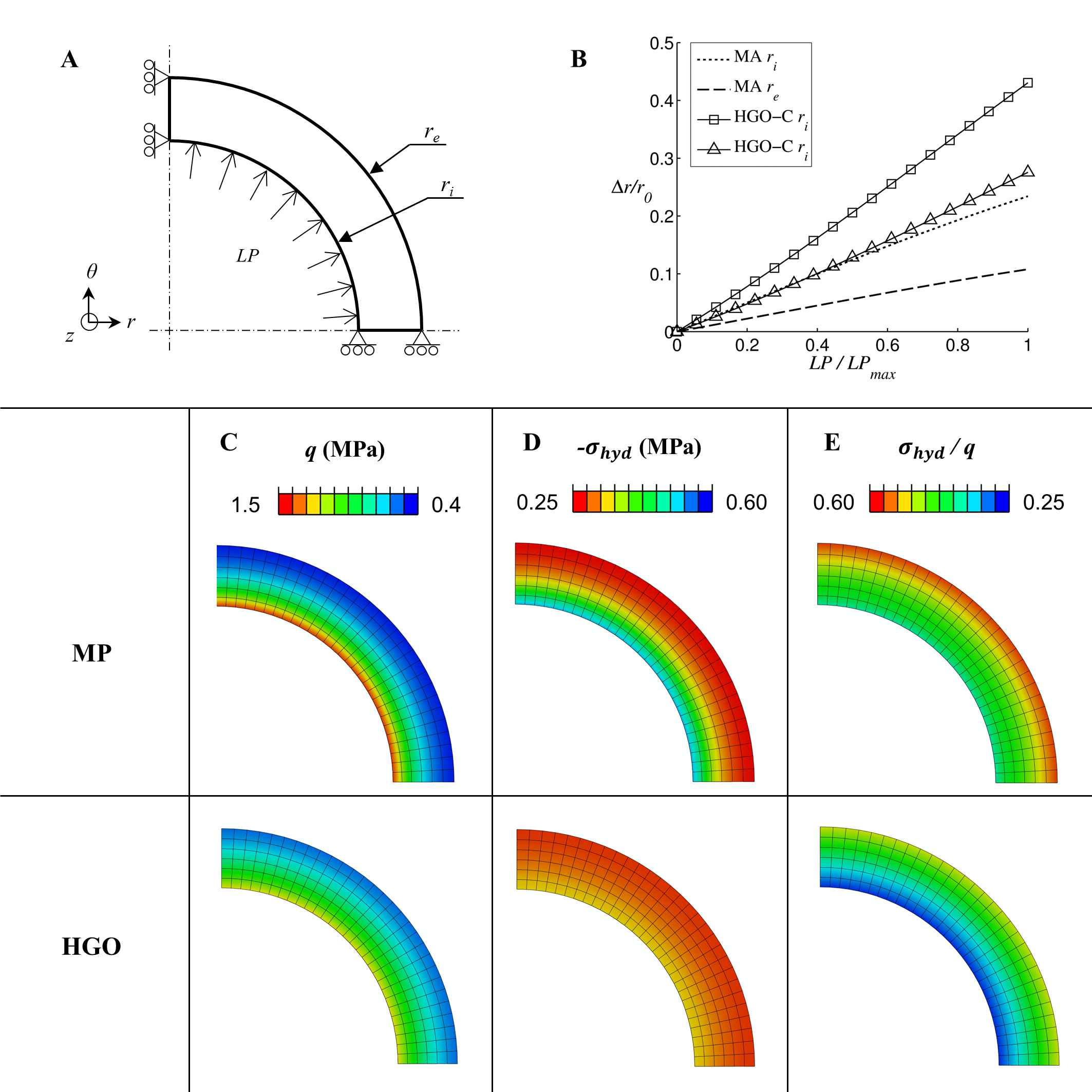}
 \caption{\textbf{ A)} Schematic illustrating the geometry, lines of symmetry and boundary conditions for modelling the inflation of an artery under a lumen pressure $LP$. \textbf{B)} Prediction of the internal ($r_{\mathrm{i}}$) and external ($r_{\mathrm{e}}$) radial strain $\Delta r/r_0=(r-r_0)/r_0$ in the artery under a normalized lumen pressure $LP/LP_{\mathrm{max}}$ for the HGO-C and MA models.  Panels \textbf{C)}, \textbf{D)} and \textbf{E)} are contour plots illustrating the von Mises ($q$), pressure ($-\sigma_{\text{hyd}}$) and triaxiality ($\sigma_{\text{hyd}}/q$) stresses, respectively,  in the artery wall for the HGO-C and MA models.}
 \label{fig:artery}
\end{figure}

First we simulate the deformation of an artery under a lumen pressure ($LP$).  A schematic of a quarter artery is shown in Figure \ref{fig:artery}A.
The vessel has an internal radius $r_{\mathrm{i}}$ of 0.6\,mm and an external radius $r_{\mathrm{e}}$ of 0.9\,mm.
The length of the artery in the $z$-direction is 0.3 mm  with both ends constrained in the $z$-direction.

We model the wall as a homogeneous material with two families of fibres lying locally in the $(\theta,z)$ plane, where $(r,\theta,z)$ are cylindrical polar coordinates.
The fibre families are symmetric with respect to the circumferential direction and oriented at $\pm 50^\circ$ measured from the circumferential direction.
For the fibres, the material constants are $k_1= 1$\,MPa and $k_2 = 2$, and for the neo-Hookean matrix, they are $\mu_{0}= 0.03$\,MPa, $\kappa_{0} = 1$\,MPa, resulting in a slightly incompressible material (corresponding to a Poisson ratio of 0.485).  A mesh sensitivity study confirms a converged solution for a model using a total of 1,044 eight-noded full-integration hexahedral elements.

The (dimensionless) changes in the internal and external radii $\Delta r/r_0$ as functions of increasing dimensionless lumen pressure $LP/LP_{\mathrm{max}}$ are plotted in Figure \ref{fig:artery}B.  They reveal that the HGO-C model predicts a far more compliant artery than the MA model.

Notable differences in the arterial wall stress state arise between the HGO-C and MA models.  Figures \ref{fig:artery}C, D and E present the von Mises stress, pressure stress and triaxiality, respectively, in the arterial wall.  The magnitude and gradient through the wall thickness of both the von Mises stress and pressure stress differ significantly between the HGO-C and MA models.
This contrast is further highlighted by the differing distributions of triaxiality for both models, confirming a fundamental difference in the multi-axial  stress state computed for the two models.

\subsection{ Stent deployment in an artery}


The final case study examines the deployment of a stainless steel stent in a straight artery.
Nowadays most medical device regulatory bodies insist on computational analysis of stents \citep{FDA2010} as part of their approval process.
Here we demonstrate that the correct implementation of the constitutive model for a slightly compressible arterial wall is critical for the computational assessment of stent performance.

We use a generic closed-cell stent geometry \citep{Conway2012} with an undeformed radius of 0.575\,mm.
It is made of biomedical grade stainless steel alloy 316L with Young's modulus of 200\,GPa and Poisson's ratio 0.3 in the elastic domain.
We model plasticity using isotropic hardening $J_2-$plasticity with a yield stress of 264\,MPa and ultimate tensile strength of 584\,MPa at a plastic log strain of 0.274 \citep{McGarry2007}.
We mesh the stent geometry with 22,104 reduced integration hexahedral elements.
We model a balloon using membrane elements, with frictionless contact between the membrane elements and the internal surface of the stent.
Finally, we simulate the balloon deployment by imposing radial displacement boundary conditions on the membrane elements.

For the artery, we take a single layer with two families of fibres symmetrically disposed in the $(\theta,z)$ plane.
The fibres are oriented at $\pm 50^\circ$ to the circumferential direction and material constants and vessel dimensions are the same as those used in Section \ref{pressure}.
Here the FE mesh consists of 78,100 full integration hexahedral elements; a high mesh density is required due to the complex contact between the stent and the artery during deployment.

\begin{figure}[!t]
\centering{
\includegraphics*[width=0.8\linewidth]{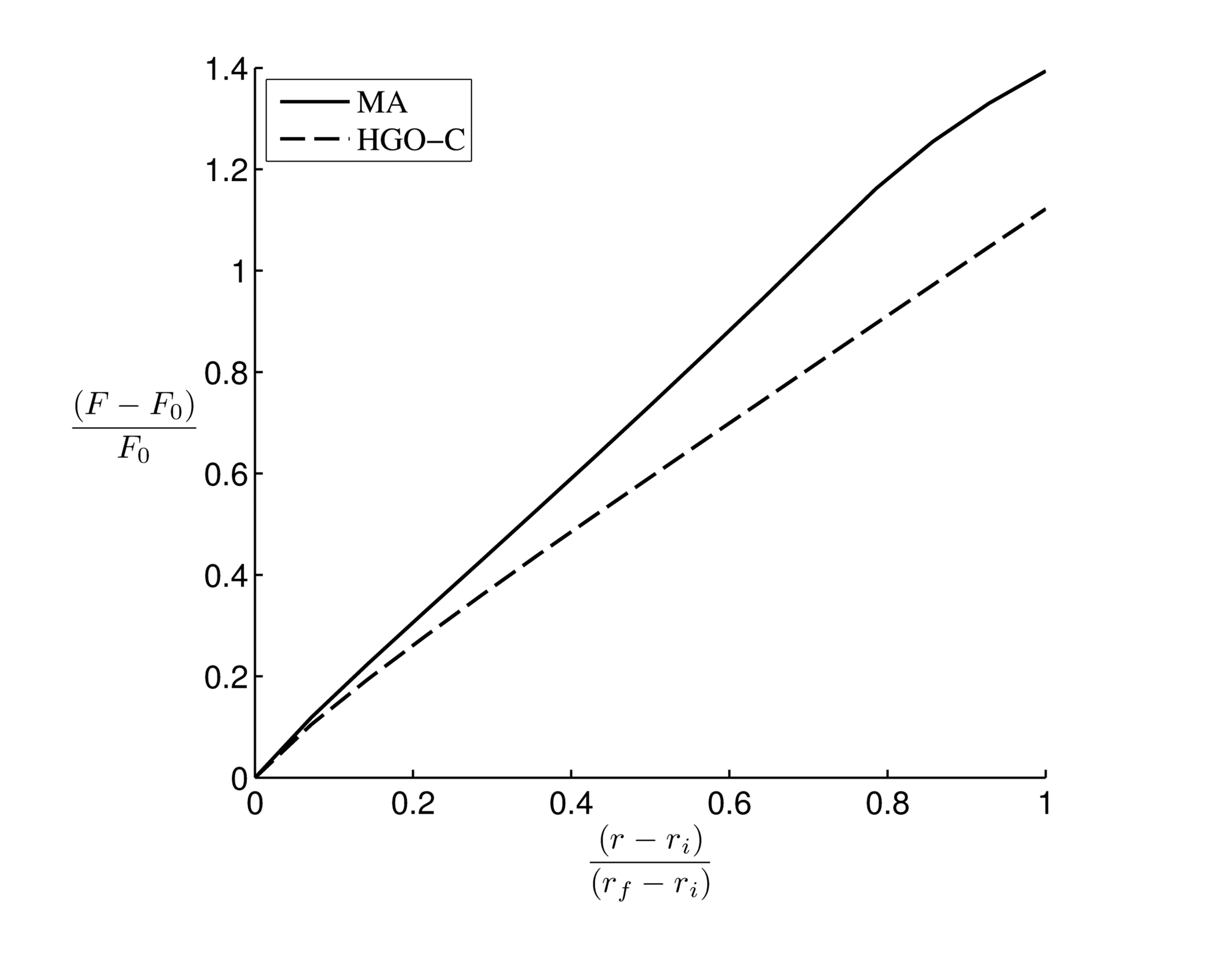}}
 \caption{Plot of the dimensionless radial force $(F-F_0)/F_0$ required to deploy a stent in an artery with increasing stent radial expansion.
Radial force is normalized by the radial force at the point immediately before contact with the artery ($F_0$).
The radial expansion is normalized using the initial undeformed internal radius ($r_{\mathrm{i}}$) and the final fully deployed internal radius ($r_{\mathrm{f}}$).
Note that the HGO-C model predicts a more compliant artery than the MA model.}
 \label{fig:force}
\end{figure}

``Radial stiffness'', the net radial force required to open a stent, is a commonly cited measure of stent performance \citep{FDA2010}.
Figure \ref{fig:force} presents plots of the predicted net radial force as a function of radial expansion for the HGO-C and MA models.
The predicted radial force required to expand the stent to the final diameter is significantly lower for the HGO-C model than for the MA model.
This result correlates with the previous finding in Section \ref{pressure} that the HGO-C model underestimates the arterial compliance, with significant implications for design and assessment of stents.

\begin{figure}[!t]
\centering
 \includegraphics*[width=1\linewidth, keepaspectratio=false]{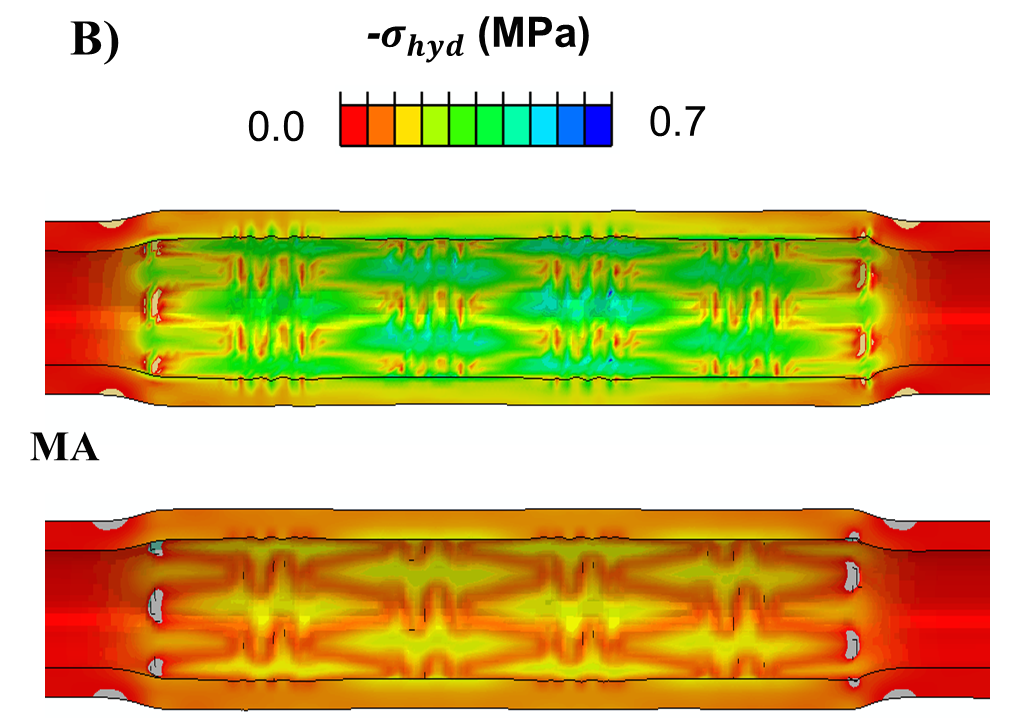}
\caption{Contour plots illustrating differences in the stresses computed for the HGO-C and MA models after stent deployment. \textbf{A)} von Mises stress \textit{q}, \textbf{B)} pressure stress -\textit{$\sigma_{\text{hyd}}$}, \textbf{C)} triaxiality, \textbf{D)} ratio of axial stress to the circumferential stress $\sigma_{zz}/\sigma_{\theta\theta}$.}
\label{fig:stent}
\end{figure}

Figure \ref{fig:stent} illustrates the notable differences that appear in the artery stress state between the HGO-C and MA models.
Again, higher values of von Mises stress (Figure \ref{fig:stent}A) and pressure stress (Figure \ref{fig:stent}B)  are computed for the MA model.
Both the triaxiality (Figure \ref{fig:stent}C) and the ratio of axial to circumferential stress (the stress ratio in the plane of the fibres) (Figure \ref{fig:stent}D) confirm that the nature of the computed multi-axial stress state is significantly different between the MA and HGO-C models.

\begin{figure}[h!]
 \includegraphics*[width=\linewidth]{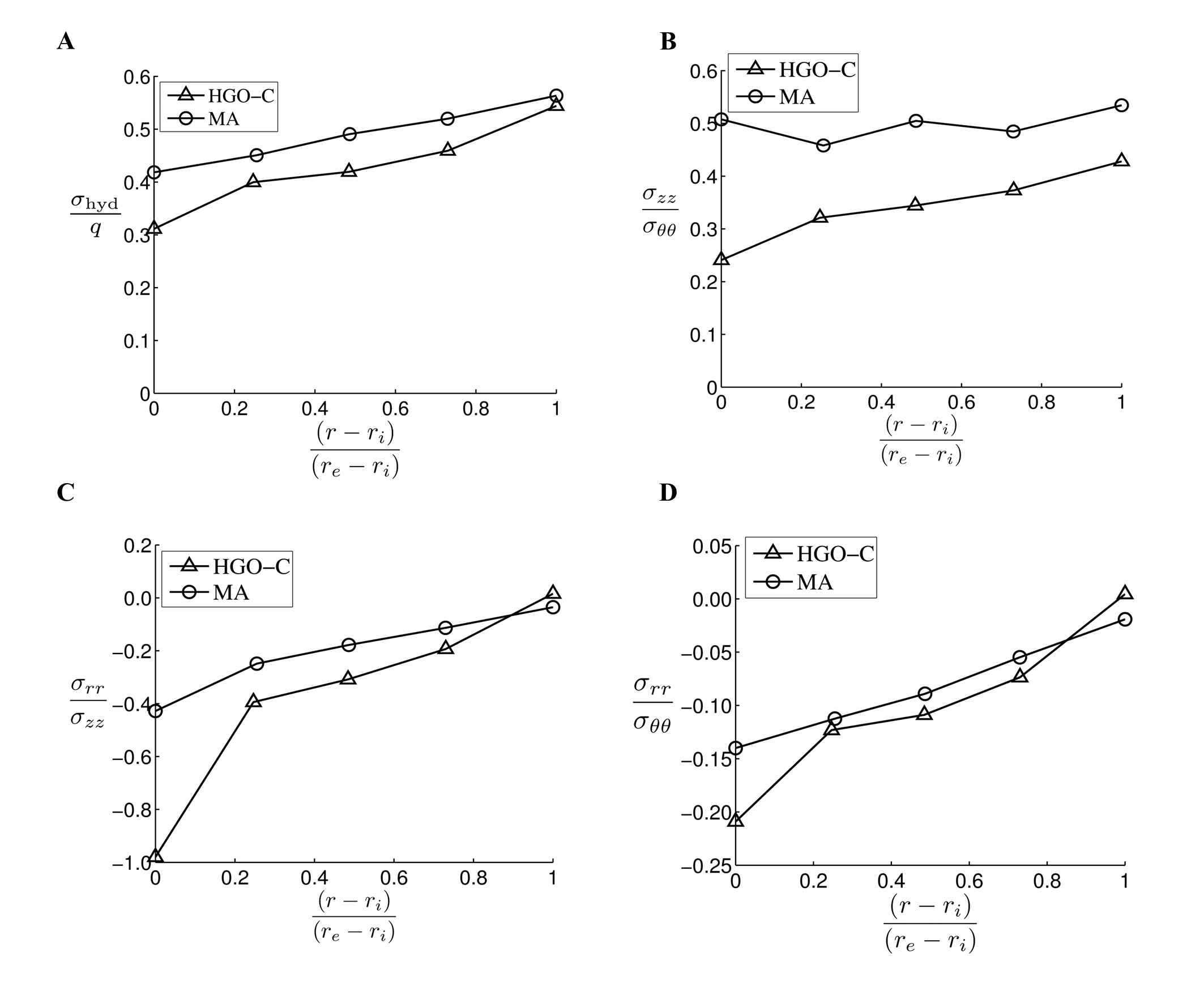}
 \caption{Stress measures computed through the arterial wall from the internal ($r_{\mathrm{i}}$) to external radius ($r_{\mathrm{e}}$) at full deployment of the stent for the HGO-C and MA models. \textbf{A)} Triaxiality ratio $\sigma_{\text{hyd}}/q$ of the pressure stress to von Mises stress. \textbf{B)} Ratio $\sigma_{zz}/\sigma_{\theta\theta}$ of the axial to circumferential stress. \textbf{C)} Ratio $\sigma_{rr}/\sigma_{zz}$ of the radial to axial stress. \textbf{D)} Ratio $\sigma_{rr}/\sigma_{\theta\theta}$ of the radial to circumferential stress.}
 \label{fig:wall}
\end{figure}

A detailed examination of the stress state through the thickness (radial direction) of the artery wall is presented in Figure \ref{fig:wall}.
A comparison between HGO-C and MA simulations in terms of the ratios of the Cauchy stress components emphasizes further the fundamentally different stresses throughout the entire artery wall thickness. It is not merely that the MA model calculates a different magnitude of stress, rather the multi-axiality of the stress state has been altered.

\section{Concluding remarks}
\label{conclusion}


The original HGO model (\cite{Holzapfel2000}) is intended for modelling of incompressible anisotropic materials. A compressible form (HGO-C model) is widely used whereby the anisotropic part of $\Psi$ is expressed in terms of isochoric invariants. Here we demonstrate that this formulation does not correctly model compressible anisotropic material behaviour. The anisotropic component of the model is insensitive to volumetric deformation due to the use of isochoric anisotropic invariants. This explains the anomolous finite element simulations reported in \cite{Vergori2013}, whereby a slightly compressible HGO-C sphere was observed to deform into a larger sphere under tensile hydrostatic loading instead of the ellipsoid which would be expected for an anisotropic material. In order to achieve correct anisotropic compressible hyperelastic material behaviour we present and implement a modified (MA) model whereby the anisotropic part of the strain energy density is a function of the total form of the anisotropic invariants, so that a volumetric anisotropic contribution is represented. This modified model correctly predicts that a sphere will deform into an ellipsoid under tensile hydrostatic loading.   

In the case of (plane strain) pure shear, a kinematically enforced isochoric deformation, we have shown that a correct stress state is computed for the MA model, whereas the HGO-C model yields incorrect results. Correct results are obtained for the HGO-C model only when incompressibility is effectively enforced via the use of a large bulk modulus, which acts as a Lagrange multiplier in the volumetric contribution to the isotropic terms (in this case HGO-C model is effectively the same as the original incompressible HGO model). In the case of a nearly incompressible material (with Poisson's ratio = 0.490, for example) we have shown that the in-plane stress components computed by the HGO-C model are reduced by an order of magnitude. Bulk modulus  sensitivity has been pointed out for isotropic models by \cite{Gent2007} and \cite{Destrade2012}, and for the HGO-C model by \cite{NiAnnaidh2013b}. Here, we have demonstrated that a ratio of bulk to shear modulus of $\kappa_{0}/\mu_{0}=2 \times 10^{6}$ (equivalent to a Poisson's ratio of 0.49999975) is required to compute correct results for the HGO-C model. By contrast, the MA model is highly robust with correct results being computed for all levels of material compressibility during kinematically prescribed isochoric deformations. 

From the view-point of general finite element implementation, red the requirement of perfect incompressibility (as in the case of a HGO material) can introduce numerical problems requiring the use of selective reduced integration and mixed finite elements to avoid mesh locking and hybrid elements to avoid ill-conditioned stiffness matrices. Furthermore, due to the complex contact conditions in the simulation of balloon angioplasty (both between the balloon and the stent, and between the stent and the artery), explicit Finite Element solution schemes are generally required. However, Abaqus/Explicit for example has no mechanism for imposing an incompressibility constraint and assumes by default that $\kappa_{0}/\mu_{0}=20$ ($\nu= 0.475$). A value of $\kappa_{0}/\mu_{0}>100 $ ($\nu= 0.495$) is found to introduce high frequency noise into the explicit solution. We have demonstrated that the HGO-C model should never be used for compressible or slightly compressible materials. Instead, due to its robustness, we recommend that the MA model is used in FE implementations because (i) it accurately models compressible anisotropic materials, and (ii) if material incompressibility is desired but can only be approximated numerically (e.g., Abaqus/Explicit) the MA model will still compute a correct stress state.

A paper by \cite{Sansour2008} outlined the potential problems associated with splitting the free energy for anisotropic hyperelasticity into volumetric and isochoric contributions; see also \cite{Federico2010} for a related discussion. A study of the HGO-C model by \cite{Helfenstein2010} considered the specific case of uniaxial stress with one family of fibres aligned in the loading, and suggested that the use of the `total' anisotropic invariant $I_i$ is appropriate.  The current paper demonstrates the importance of a volumetric anisotropic contribution for compressible materials, highlighting the extensive range of non-physical behaviour that may emerge in the simulation of nearly incompressible materials if the HGO-C model is used instead of the MA model. Examples including the Finite Element analysis of artery inflation due to increasing lumen pressure and stent deployment. Assuming nearly incompressible behaviour ($\nu= 0.485$) the HGO-C model is found to significantly underpredict artery compliance, with important implications for simulation and the design of stents \citep{FDA2010}. We have shown that the multiaxial stress state in an artery wall is significantly different for the HGO-C and MA models. Arterial wall stress is thought to play an important role in in-stent restenosis (neo-intimal hyperplasia) \citep{Thury2002, Wentzel2003}.
Therefore, a predictive model for the assessment of the restenosis risk of a stent design must include an appropriate multiaxial implementation of the artery constitutive law. 

\section*{\textbf{Acknowledgements}}
\noindent DRN and ALG wish to acknowledge the receipt of their PhD scholarships from the Irish Research Council. MD and RWO wish to thank the Royal Society for awarding them an International Joint Project. This research was also funded under Science Foundation Ireland project SFI-12/IP/1723.



\appendix



\section{Consistent Tangent Matrix}


To write a UMAT, we need provide the Consistent Tangent Matrix (CTM) of the chosen model. When expressed in terms of Cauchy stress  the CTM given in \cite{Abaqus2010} may be written as
\begin{equation}
\mathcal{C}_{ijkl} = \sigma_{ij} \delta_{kl} +\frac{1}{2}\left (\frac{\partial \sigma_{ij}}{\partial F_{k \alpha}} F_{l\alpha}  + \frac{\partial \sigma_{ij}}{\partial F_{l\alpha}} F_{k\alpha} \right ),
\end{equation}
which has both the $i\leftrightarrow j$ and $k\leftrightarrow l$ minor symmetries. 

The CTM may estimated using either numerical techniques or an analytical solution. Here we first describe a numerical technique for estimation of the CTM. We then present the analytical solution for the MA and HGO-C CTM. 


\subsection*{Numerical Approximation of the CTM}

The CTM may be approximated numerically (\cite{Sun2008}), and a short overview is presented here. This numerical approximation is based on a linearised incremental form of the Jaumann rate of the Kirchhoff stress:
\begin{equation}\label{eq_jaumann}
\Delta \vec{\tau} - \Delta \vec{W} \vec{\tau} - \vec{\tau} \Delta \vec{W}^{T} = \mathcal{C}:\Delta \vec{D}, 
\end{equation} 
where $\vec{\tau}$ is the Kirchhoff stress, $\Delta \vec{\tau}$ is the Kirchhoff stress rate, $\Delta \vec{D}$ the rate-of-deformation tensor and $\Delta \vec{W}$ the spin tensor are the symmetric and anti-symmetric parts of the spatial velocity gradient $\Delta \vec{L}$ (where $\Delta \vec{L}=\Delta \vec{F} \vec{F}^{-1}$), and $\mathcal{C}$ is the CTM. 

To obtain an approximation for each of components of the CTM, a small perturbation is applied to \eqref{eq_jaumann} through $\Delta \vec{D}$. This is achieved by perturbing the deformation gradient six times, once for each of the independent components of $\Delta \vec{D}$, using
\begin{equation}
\Delta \vec{F}^{(ij)} = \dfrac{\epsilon}{2}(e_i \otimes e_j \vec{F} + e_j \otimes e_i \vec{F}),
\end{equation}
where $\epsilon$ is a perturbation parameter, $e_i$ is the basis vector in the spatial description, $(ij)$ denotes the independent component being perturbed. 

The `total' perturbed deformation gradient is given by $\vec{\hat{F}}^{(ij)} = \Delta \vec{F}^{(ij)} + \vec{F}$. The Kirchhoff stress is then calculated using this perturbed deformation gradient ($\vec{\tau}(\vec{\hat{F}}^{ij})$). The CTM is approximated using
\begin{equation}\label{eq_CTM_approx}
\mathcal{C}^{(ij)} \approx \dfrac{1}{J\epsilon}(\vec{\tau}(\vec{\hat{F}}^{(ij)}) - \vec{\tau}(\vec{F})),
\end{equation}
where $J$ is the determinant of the deformation gradient. Each perturbation of \eqref{eq_CTM_approx} will produce six independent components. This is performed six times for each independent $(ij)$, giving the required $6 \times 6$ CTM matrix.


\subsection*{Analytical solutions for the MA and HGO-C CTM}

Here we present an analytical solution for the CTM for the MA and HGO models. For convenience we give the volumetric, isotropic and anisotropic contibutions separately.

For the MA model the stress is given by equations~\eqref{devia} and \eqref{GrindEQ__stressmp_}. We can calculate $\mathcal{C}_{ijkl}$ from
\begin{align}
\label{eqn:CTMvol}
& (\vec{\sigma}_\text{vol})_{ij} \delta_{kl} + \frac{\partial  (\vec{\sigma}_\text{vol})_{ij}  }{\partial F_{k \alpha}}F_{l \alpha}  = \kappa_{0} (2 J-1) \delta _{i j} \delta _{k l},
\\
\label{eqn:CTMiso}
&   (\vec{\overline{\sigma}}_\text{iso})_{ij}  \delta_{kl} + \frac{\partial  (\vec{\overline{\sigma}}_\text{iso})_{ij} }{\partial F_{k \alpha}}F_{l \alpha} =  \mu_{0} J^{-1} \left( \overline{B}_{j l} \delta _{i k}+\overline{B}_{i l} \delta_{j k}-\tfrac{2}{3} \overline{B}_{i j} \delta _{k l} -\tfrac{2}{3} \overline{B}_{k l} \delta _{i j} +\tfrac{2}{9} \overline{I}_1 \delta _{i j} \delta _{k l}\right),
   \\
 &  (\vec{\sigma}_\text{aniso}) _{ij} \delta_{kl} +  \frac{\partial  (\vec{\sigma}_\text{aniso}) _{ij} }{\partial F_{k \alpha}}F_{l \alpha} = 2k_1J^{-1}\sum_{n=4,6} ( I_n -1)\exp[k_2(I_n-1)^2]  \left( a_{nj} a_{nl} \delta _{i k}+a_{ni} a_{nl} \delta _{j k}\right)
 \notag \\
 &\qquad\qquad\qquad   +4k_1 J^{-1}\sum_{n=4,6} [ 2 (I_n-1)^2 k_2+1] \exp[k_2(I_n-1)^2] a_{ni} a_{nj} a_{nk} a_{nl},
\end{align}
where we have used $a_{ni},\,n=4,6,\,i=1,2,3$, is the $i$th component of $\vec{a}_n = \vec{ F}\vec{a}_{0n}$. 
\\
\\
For the HGO-C model the stress is given by equations~\eqref{devia} and~\eqref{GrindEQ__strhgo_}.
Once again the isotropic contributions to $\mathcal{C}_{ijkl}$ are given by equations~\eqref{eqn:CTMvol} and~\eqref{eqn:CTMiso}. The anisotropic contribution to $\mathcal{C}_{ijkl}$ for the HGO-C model is given as:
\begin{align}
 &  (\vec{\overline{\sigma}}_\text{aniso}) _{ij} \delta_{kl} + \frac{\partial  (\vec{\overline{\sigma}}_\text{aniso}) _{ij} }{\partial F_{k \alpha}}F_{l \alpha} =
   4k_1J^{-1}\sum_{n=4,6}[1+2 k_2 \left(\overline{I}_n-1\right)^2]\exp[k_2 \left(\overline{I}_n-1\right)^2] \notag \\
  &\hspace*{2.5in} \times\left( \overline{a}_{ni} \overline{a}_{nj} -\tfrac{1}{3}\overline{I}_n\delta_{ij}\right)
   \left( \overline{a}_{nk} \overline{a}_{nl}  -\tfrac{1}{3}\overline{I}_n\delta_{kl}\right)
  \notag \\
  &
  \qquad\qquad  +2k_1J^{-1}\sum_{n=4,6}(\overline{I}_n-1) \exp[k_2 \left(\overline{I}_n-1\right)^2]\left(\delta _{ik}\overline{a}_{nj} \overline{a}_{nl}+\delta _{jk}\overline{a}_{ni} \overline{a}_{nl}\right. \notag \\
  &\left.\hspace*{2.5in}-\tfrac{2}{3}\delta _{kl}\overline{a}_{ni} \overline{a}_{nj}
    -\tfrac{2}{3}\delta _{ij}\overline{a}_{nk} \overline{a}_{nl}+\tfrac{2}{9}\overline{I}_n\delta_{ij}\delta_{kl}\right),
\end{align}
where $\overline{a}_{ni}$ is the $i$th component of $\overline{\vec{a}}_n = \vec{\overline{F}}\vec{a}_{0n}$.

\end{document}